\documentstyle[sprocl,psfig]{article}

\def\ep{\epsilon}

\def\Journal#1#2#3#4{{#1} {\bf #2}, #3 (#4)}

\def\NPB{{\em Nucl. Phys.} B}
\def\PLB{{\em Phys. Lett.}  B}
\def\PRL{\em Phys. Rev. Lett.}
\def\PRD{{\em Phys. Rev.} D}
\def\ZPC{{\em Z. Phys.} C}

\def\be{\begin{equation}}
\def\ee{\end{equation}}
\def\bea{\begin{eqnarray}}
\def\eea{\end{eqnarray}}

\begin{document}
\vspace*{-10mm}
\begin{flushright}
BNL-HET/98-22\\
TTP 98--23\\
hep-ph/9806258\\[3mm]
\end{flushright}

\title{Perturbative QCD calculations with heavy quarks}

\author{Andrzej Czarnecki}
\address{Physics Department, Brookhaven National Laboratory,\\
Upton, NY 11973}

\author{Kirill Melnikov}
\address{Institut f\"ur Theoretische Teilchenphysik, 
Universit\"at Karlsruhe,\\
D-76128 Karlsruhe, Germany}


\maketitle
\abstracts{ 
We review recent calculations of second order corrections to heavy
quark decays.  Techniques developed in this context are applicable to
many processes involving heavy charged particles, for example the
nonabelian dipole radiation.  Calculations involving relativistic
radiating particles are also discussed.
}

\section{Introduction}
Description of the higher order QCD effects in heavy quark processes
is important in view of the improving experimental precision.  Ongoing
efforts to determine the CKM matrix elements motivate a thorough
theoretical description of $B$ meson decays (for a review see
\cite{BSU}).  These studies are made difficult by the complexity of
the initial state, however even their conceptually simplest part ---
perturbative ${\cal O}(\alpha_s^2)$ QCD corrections to a free quark
decay --- were not known until recently.  Their expected size
was controversial and various authors used to assign widely different
theoretical uncertainties in predictions for the 
semileptonic $B \to X_c l \nu_l$ decays.

In this talk we present a summary of our recent results on the second
order corrections to charged fermion decays.  We also discuss other
applications of the techniques developed in those calculations. 

An exact calculation of the second order corrections to a quark (or,
more generally, any charged particle) decay is presently not possible
for technical reasons.  On the other hand, none of the standard
methods of approximate calculations (for a short review see \cite{VS})
seems to be useful for those problems, mainly because it is not clear
what could play a role of a small parameter.  Here we will demonstrate
how one can employ expansions in rather large parameters to obtain
good accuracy of approximate results.  With the example of the
one--loop corrections to the top quark decay we will also show that in
some cases such expansions can also give exact analytical results.

\section{Computational techniques}
As an example we first consider the semileptonic decay $b \to c l
\nu_l$.  Its total width can be expressed as an integral over the
invariant mass of the leptons
\be
\Gamma = \int \limits_{0}^{(m_b-m_c)^2} 
{\rm d}q^2 \frac {{\rm d}\Gamma}{{\rm d}q^2}.
\ee
A good estimate of ${\cal O}(\alpha _s^2)$ effects
in $\Gamma$ can be obtained by computing  ${\cal
O}(\alpha_s^2)$ corrections to ${\rm d}\Gamma/{\rm d}q^2$ at several
values of $q^2$.  For the sake of clarity we focus on the point
$q^2=0$; similar methods are applicable in a more general context.

To order $\alpha_s^2$ one has to consider three sources of
corrections.  Namely, there are virtual corrections, the one-loop
corrections to a single gluon emission, and the radiation of two
gluons. Consider the situation when the masses of the $c$ quark and
the $b$ quark are very close to each other. In this case, the
kinematics of the process simplifies considerably -- the final $c$
quark is produced almost at rest and moves slowly; as a result, the
radiation of real gluons is suppressed.  The calculations in the limit
$m_b-m_c \ll m_b$ can be considerably simplified.  It is, however,
more important that these calculations can be formulated in an
algorithmic form --- which shifts the burden of the calculations to
the computer and makes it possible to obtain many terms in the
expansion in $\delta = 1-m_c/m_b$.

Consider for example the two-loop virtual corrections.  In a general
case, they are a function of kinematic invariants like the momentum
transfer to the leptons, and two different masses. In the limit $m_c
\to m_b$, the virtual corrections simplify --- they become dependent
on a single parameter $m_b$ only.  To account for the difference
$m_b-m_c$ we expand the diagrams in $\delta$.  The coefficients of the
expansion are single scale Feynman diagrams.  In higher order terms in
this expansion the propagators in those diagrams can have high powers.
Integration by parts techniques \cite{che81} provide algebraic
relations among these integrals.  A solution of those relations allows
us to reduce any diagram to a set of ``basic'' integrals.~\cite{bro91a}

Consider now the one-loop corrections to the real gluon emission. We
essentially follow the same strategy as above. The peculiarity of this
contribution is that the Taylor expansion in $\delta$ is not
sufficient and one has to resort to the so-called eikonal
expansion.~\cite{eiko,maxtech} For the radiation of two gluons, a
special parameterization of the phase space allows us to obtain a
systematic expansion in $\delta$.

\section{Decay $b \to c l \nu_l$ and other applications with heavy
quarks}
The techniques outlined in the preceding Section enabled us to 
calculate the ${\cal O}(\alpha_s^2)$ corrections to the
differential width ${\rm d}\Gamma(b \to c l \nu_l)/{\rm d}q^2$ for
three values of the lepton invariant
mass~\cite{Czarnecki:1997hc,onehalf,zeroan}: $q^2=0$, $q^2=m_c^2$, and
$q^2=(m_b-m_c)^2$.  In this last point the BLM \cite{BLM}
corrections were known
previously.~\cite{Neubert95beta} 
Using these results to fit the $q^2$
dependence of the second order correction, we estimated~\cite{onehalf}
the second order QCD corrections to the total decay width of the
$b$ quark $b \to c l \nu_l$:
\be
\Gamma_{\rm sl} = {G_F^2 m_b^5\over 192\pi^3}|V_{cb}|^2 F \left
( \frac {m_c^2}{m_b^2} \right ) \left[ 1 -1.67 \frac {\alpha_s
(\sqrt{m_bm_c})}{\pi} -\left (9.8 - 1.4\pm 0.4\right ) \left ( \frac
{\alpha_s}{\pi} \right )^2 \right].
\label{widthpole}
\ee 
Here $F(x) = 1-8x-12x^2\ln x + 8x^3 - x^4$ and $m_b$ and $m_c$ are
the pole masses of $b$ and $c$ quarks for which we use $m_c/m_b =
0.3$.  For the sake of clarity we separated the BLM 
($-9.8$) and the non--BLM ($1.4$) parts of the second order
corrections.  We also indicated the uncertainty of our estimate of the
second order non--BLM correction.  The BLM corrections were calculated
by Luke {\it et al.} and by Ball {\it et
al.}~\cite{Luke:1995,Ball:1995} There appears to be a small
discrepancy between these results, with the latter group obtaining a
magnitude larger by 1.5\% than the former.~\cite{BBpriv} This
difference is smaller than the non-BLM error estimate and we neglect
it here.

The second order corrections in eq.~(\ref{widthpole})
appear to be rather large, particularly because of the large BLM
corrections. This is related to using the pole quark masses in the
expression for the width.~\cite{upset,BB}  It is possible to introduce
appropriate (short-distance) masses, which reduce the magnitude of the
second order perturbation corrections in (\ref{widthpole}) and improve
the convergence of the perturbation series (see \cite{onehalf} and
references therein).

From a more general perspective, the techniques described above
provide an algorithm for performing the Heavy Quark Expansion of 
Feynman diagrams to order ${\cal O}(\alpha_s^2)$.  The
applicability of these methods is therefore not limited to the total
semileptonic decay width of a heavy quark.  We now discuss one
interesting example which can be worked out using these methods.

We consider a generalization of the QED results for the dipole
radiation to the nonabelian case of QCD.  We will later comment on why
this is useful in practice.  Let us consider \cite{CzMeUr1} a
process of scattering of a color-singlet ``weak'' current $J$ with
momentum $q$ on a heavy quark $Q$ in the small velocity (SV)
kinematics.  For simplicity, the initial quark is assumed to rest.
The initial $Q$ and final $\tilde Q$ quarks can have arbitrary
masses; however, both masses must be large, so that the
nonrelativistic expansion can be applied.  The SV limit $\vec{v}=
{\vec q}/m_{\tilde Q}$, $|\vec{v}\,| \ll 1$, is kept by adjusting
$\vec{q}$.  For simplicity we consider the case of equal
masses, $m_{\tilde Q}=m_Q$, although nothing depends on this
assumption.  The current $J$ must have a non-vanishing tree level
nonrelativistic limit (e.g., $J_0^\dagger =\bar{Q} \gamma_0 \tilde Q$
or $J_S^\dagger =\bar{Q} \tilde Q$), otherwise it can be arbitrary.

Inclusive processes of the scattering of heavy quarks are described
by an appropriate structure function $W(q_0,\,\vec{q}\,)$ which is a
sum of all transition probabilities induced by $J$ into final states
with momentum $\vec{q}$ and energy $m_Q +q_0$.  The optical theorem
relates it to the discontinuity of the forward transition amplitude
$T(q_0,\,\vec{q}\,)$ at physical values of $q_0\,$:
\begin{equation}
T(q_0,\,\vec{q}\,)\;=\; \frac{i}{2m_Q}
\int\, d^4 x\: {\rm e}\,^{-iqx}\;
\langle {Q} | {T\,J(x) J^\dagger (0)} |{Q} \rangle \;,\;\;
W(q_0,\,\vec{q}\,) = 2\: \mbox {Im}\: T(q_0,\,\vec{q}\,)\;.
\label{9}
\end{equation}

The structure function $W$ takes the following form in the heavy quark
limit:
\begin{equation}
W(\omega,\,\vec{v})\;=\; N\,\delta(\omega) +
\,\frac{2\vec{v}^{\,2}}{3}\, \frac{d(\omega)}{\omega} + {\cal
O}\left(\vec{v}^{\,4}\right)\;.
\label{14}
\end{equation}
At $\vec v =0$ only the elastic peak is present.  The dipole radiation is
described by $d(\omega)$ and $N$ is a (velocity-dependent)
factor depending on the current.

In close analogy to QED, we define the dipole coupling
$\alpha_s^{(d)}$ by projecting Eq. (\ref {14}) on its second term:
\begin{equation}
C_F \frac{\alpha_s^{(d)}(\omega)}{\pi \omega}\;=\;
\lim_{\vec v \rightarrow 0} \;
\lim_{m_Q \rightarrow \infty}\,\frac{3}{2\vec{v}^{\,2}}
\frac{W(\omega,\,\vec{v})}{\int_0^{\omega}\,
W(\omega',\,\vec{v})\,{\rm d}\omega'}
\;.
\label{16}
\end{equation}
Here $\omega$ is assumed to be positive.  The denominator in the last
ratio eliminates the overall normalization of the effective
nonrelativistic current.  
The OPE and factorization of the infrared effects ensures that
$\alpha_s^{(d)}(\omega)$ is process independent.

Using the technique described in Section $2$ we obtain the dipole
coupling constant to second order in perturbation theory:
\be
\frac{\alpha_s^{(d)}(\mu)}{\pi} =
\frac{\alpha_s^{\overline{\rm MS}}(\mu)}{\pi} +
\left[\left(\frac{5}{3} - \ln{2}\right)\frac{\beta_0}{2}
- C_A \left(\frac{\pi^2}{6}-\frac{13}{12} \right)\right]
\left(\frac{\alpha_s}{\pi}\right)^2\;,
\label{eq:zwiaz}
\end{equation}
where $C_A=N_c$ for an ${\rm SU}(N_c)$ gauge group and
$\beta_0=\frac{11}{3}C_A-\frac{2}{3}N_L$ ($N_L$ is the number of flavors
contributing to the running of $\alpha_s$). 

An interesting feature of this coupling constant is that 
it determines the renormalization group evolution of the 
basic parameters of the heavy quark expansion
like $\overline{\Lambda}(\mu)$, $\mu_\pi^2(\mu)$, etc.,
in a scheme suggested in \cite{5infty}.

In perturbation theory, the knowledge of the dipole
coupling permits us to obtain perturbative expressions
for ${\overline \Lambda}(\mu)$, $\mu_\pi^2(\mu)$.
These relations can be used  to derive a relation 
\cite{CzMeUr1} between the pole mass of the quark and the so-called
low-scale kinetic mass of the heavy quark \cite{5infty} to order
${\cal O}(\alpha_s^2)$. The phenomenological
importance of the low scale running quark masses is described
in  detail in \cite{Klecture}. 

Similarly to the above calculation, it is possible to obtain
the ${\cal O}(\alpha_s^2)$ correction to the small velocity sum rules
for heavy flavor transitions, relevant for the determination of
$|V_{cb}|$ from the exclusive decays $B \to D^* l \nu_l$.~\cite{CzMeUr}

\section{Decay $t \to b W$}
The techniques described above are useful when the quark in the final
state can be considered non-relativistic.  This obviously is not the
case when the final quark is massless.  However, such applications as
the QED corrections to the muon decay, $b \to u l \nu_l$ decays, as
well as the top quark decay --- all belong to this type of decay
process.  In this section we discuss possible approaches to the
calculation of the ${\cal O}(\alpha_s^2)$ corrections to top quark
decay $t \to b W$.  We will present two alternative approaches, both
based on the introduction of an artificial small parameter.

First, one can try to apply the same strategy as indicated above for
$b \to c$ transitions to the decay $t \to b W$, if one neglects the
mass of the $W$ boson.  We note that this approximated, tested at the
${\cal O}(\alpha_s)$ corrections deviates from a complete result on
the level of $10\%$.  Then, one expands around configuration of equal
$t$ and $b$ masses, with the expansion parameter $\delta = 1 -
m_b/m_t$.  In reality, this expansion parameter is close to one,
therefore one needs to expand up to rather high powers of $\delta$.
To ${\cal O}(\alpha_s^2)$ accuracy, we obtained for the top decay
width into massless $W$ and $b$:~\cite{topCzMe}
\be
\Gamma_t = \frac {G_F m_t^3}{8 \pi \sqrt{2}} \left ( 1-0.866
\alpha_s - 1.69(5) \alpha_s^2
\right ),
\ee
where $\alpha_s$ is the $\overline {\rm MS}$ QCD coupling constant
evaluated at the scale $m_t$ and $m_t$ is the pole mass of the top
quark. More details about this calculation 
can be found in \cite{topCzMe}.

We would also like to present another approach to the same problem and
show that in some cases introduction of an artificial expansion
parameter yields exact analytical results after appropriate
resummation is performed.  Though we will discuss only ${\cal
O}(\alpha_s)$ correction to top decay in what follows, a similar
technique can be used in other applications as well.  For example, in
the case of $Z$ decay into massless quarks \cite{Czarnecki:1996ei}
this method was used to evaluate mixed electroweak-QCD corrections.
For some of the diagrams it turned out possible to sum up the
expansion exactly and obtain an analytical formula
\cite{Czarnecki:1997ac} which would have been much harder to derive
directly.

Let us briefly describe the basic idea of the method. 
Consider the self-energy of the top quark at the value 
of the incoming momenta $k^2=m^2$ smaller 
than the physical value  of the top quark mass $M$. 
If $m=M$, the imaginary part of this self-energy 
gives the decay width of the top quark.
For $m < M$ the imaginary part is still
there because of  cuts through the massless $b$ and
$W$ lines. The idea is to consider an expansion of the self-energy
diagrams in $\delta=m^2/M^2$, and 
then take the limit $\delta \to 1$ to obtain the 
result for the physical decay width of the top quark.
\begin{figure}[hbt]
\hspace*{2mm}
\begin{minipage}{16cm}
\begin{tabular}{cc}
\psfig{figure=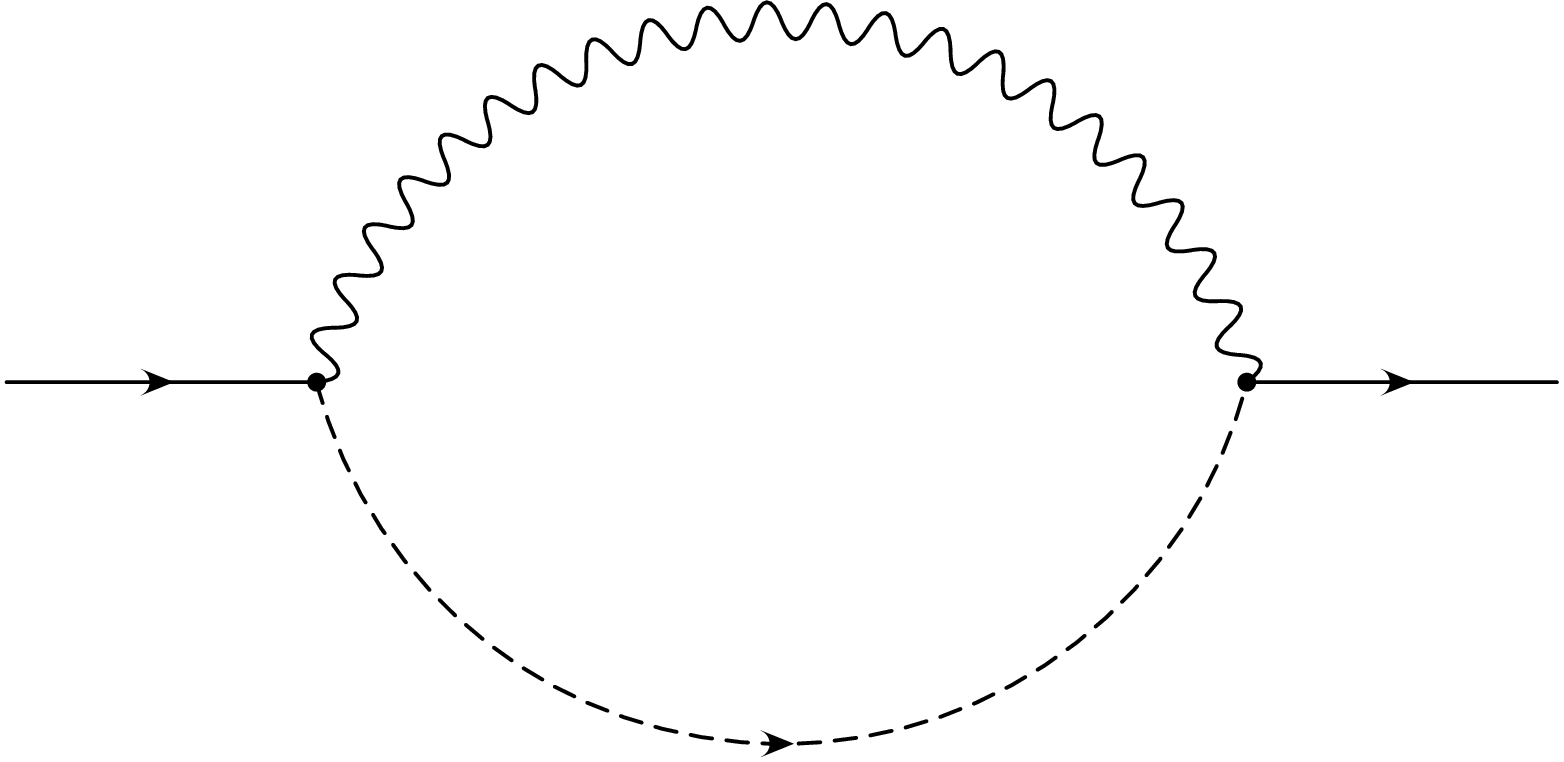,width=4.5cm,bbllx=000pt,%
bblly=390pt,bburx=630pt,bbury=550pt}  
&
\hspace*{15mm}
\psfig{figure=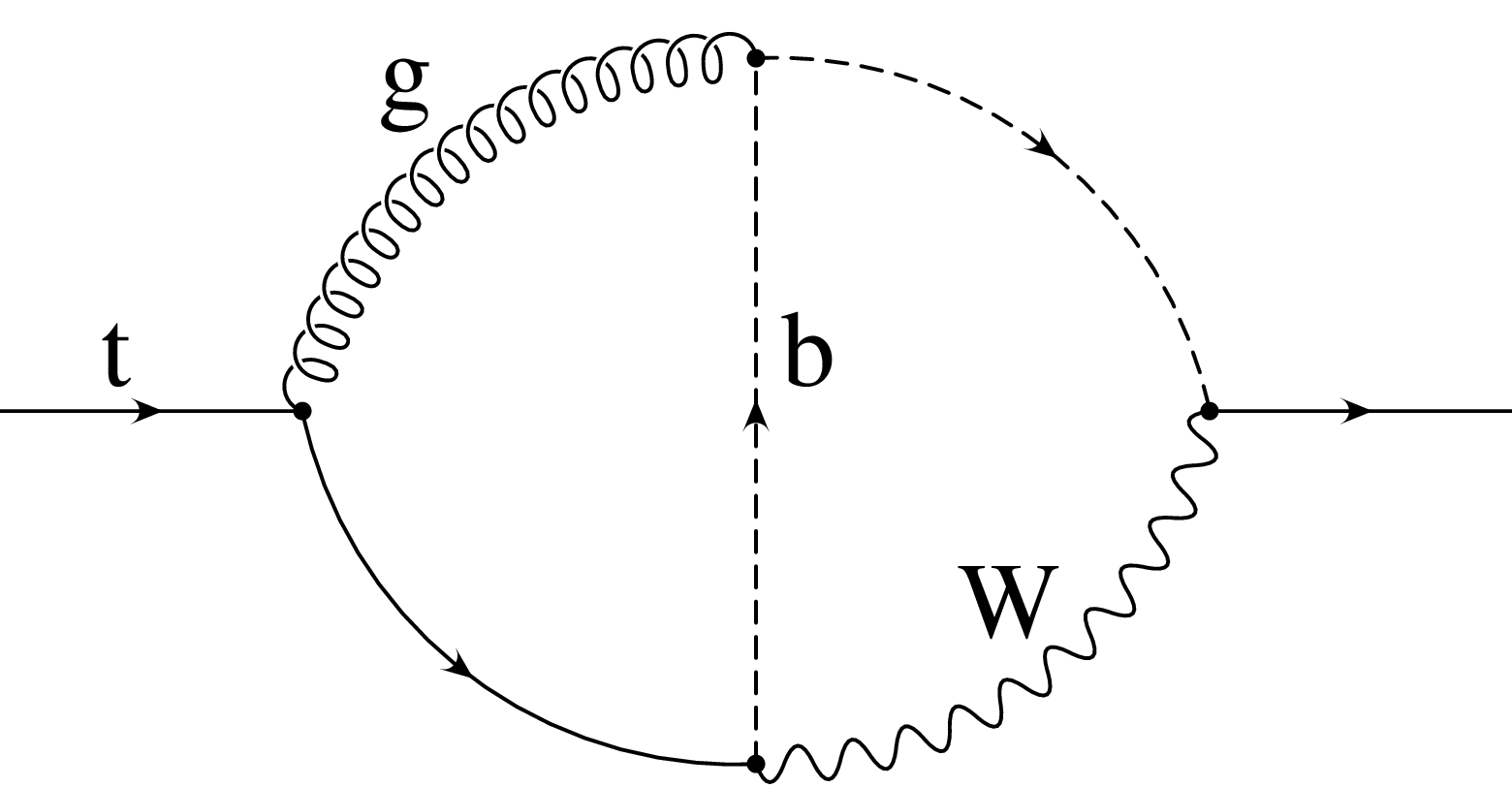,width=4.5cm,bbllx=000pt,%
bblly=390pt,bburx=630pt,bbury=550pt}  
\\[12mm]
(a) & \hspace*{14mm} (b)
\\[6mm]
\psfig{figure=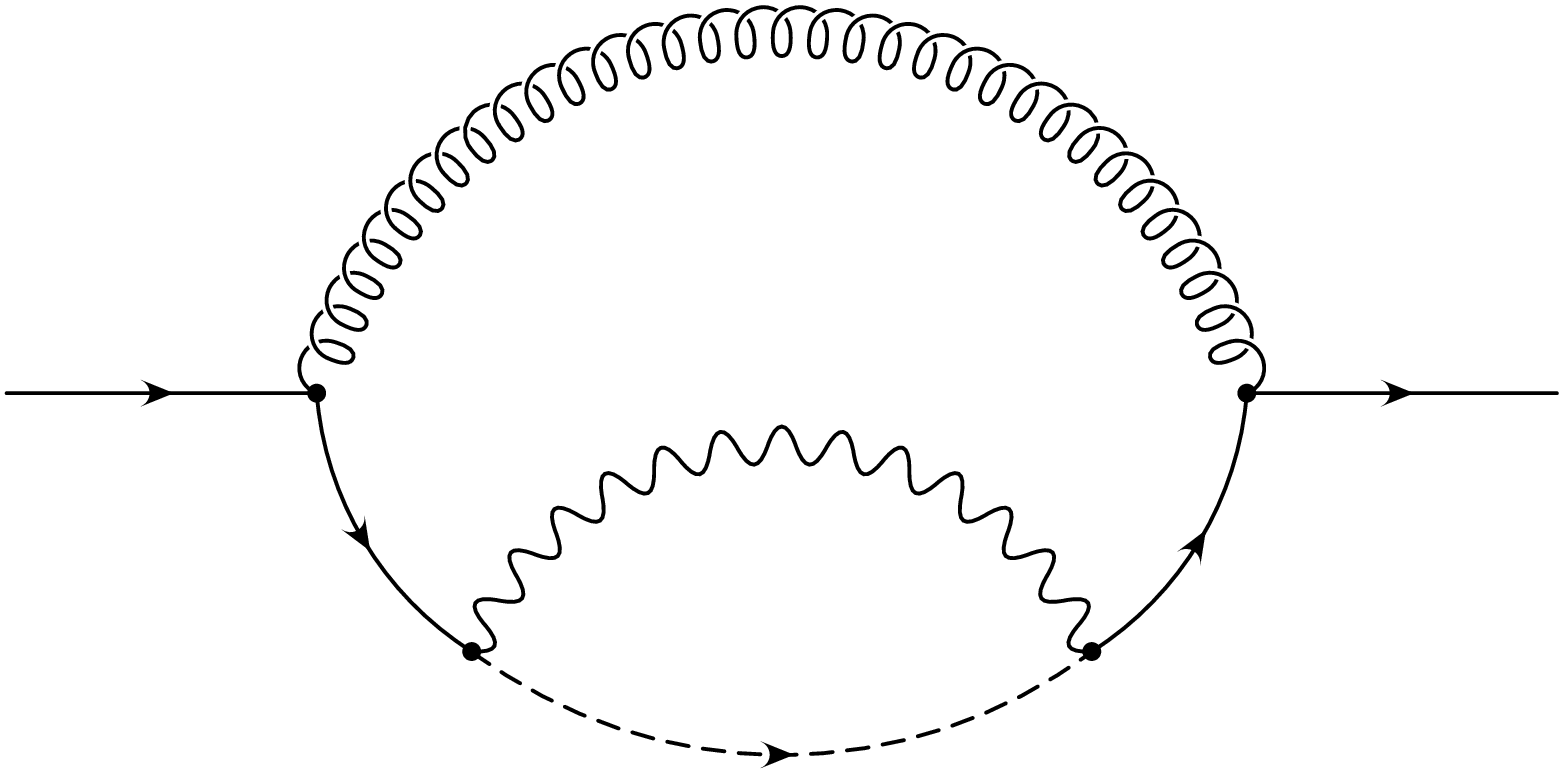,width=4.5cm,bbllx=000pt,%
bblly=390pt,bburx=630pt,bbury=550pt}  
&
\hspace*{15mm}
\psfig{figure=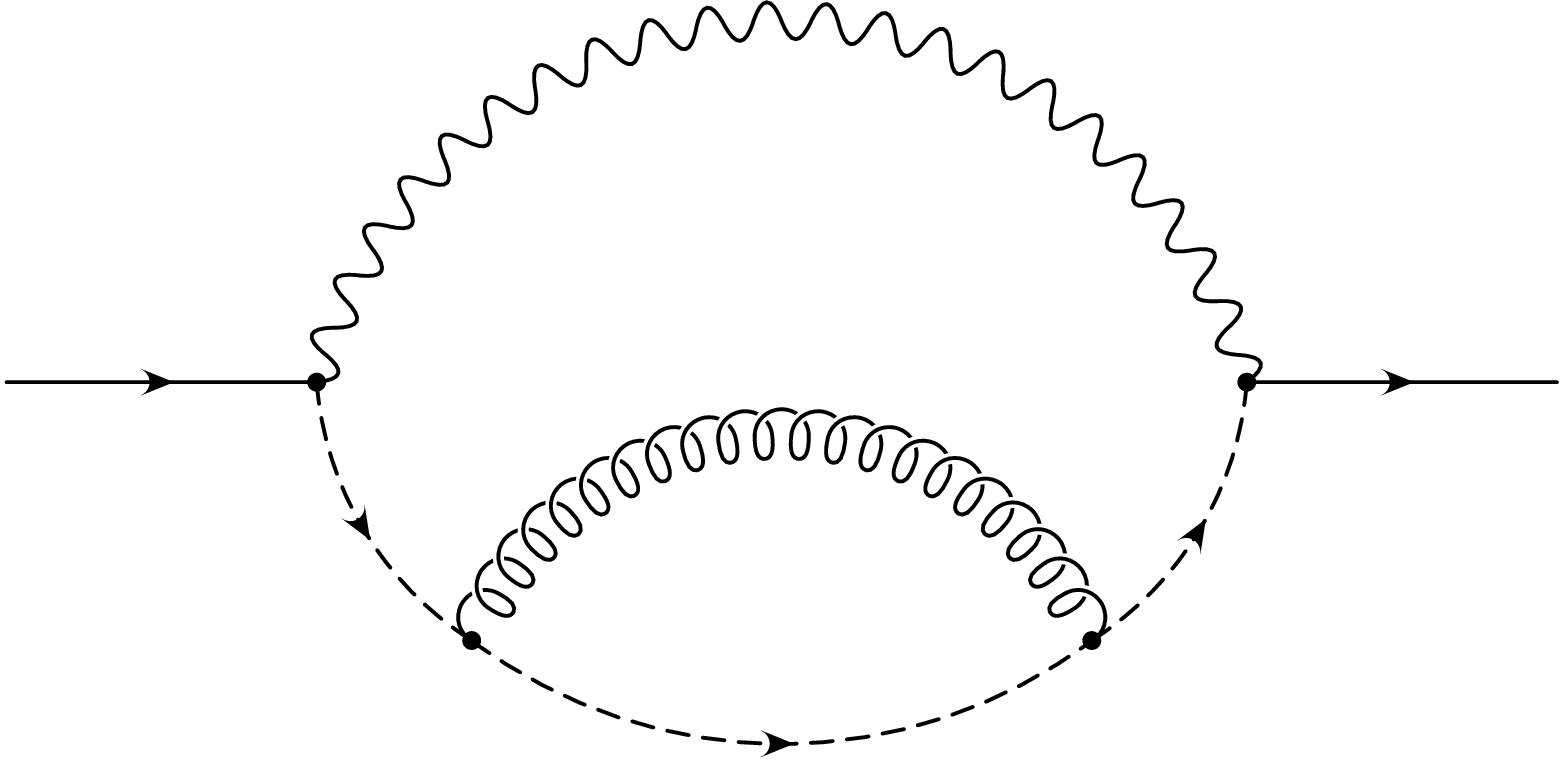,width=4.5cm,bbllx=000pt,%
bblly=390pt,bburx=630pt,bbury=550pt}  
\\[12mm]
(c) & \hspace*{14mm} (d)
\end{tabular}
\end{minipage}
\caption{Diagrams whose imaginary parts determine the top width: 
  Born level (a),   and ${\cal O}(\alpha_s)$ corrections (b,c,d).}
\label{fig:TopDecay}
\end{figure}

To order $\alpha_s$ the width of the top quark is given by the
diagrams in Fig.~\ref{fig:TopDecay}.  The limit of a very heavy top is
taken by neglecting $b$ mass and replacing the $W$ propagator by 
\be
i{p^\mu p^\nu\over p^2}{1\over M_W^2}.  
\ee
The tree level width (diagram \ref{fig:TopDecay}(a)) is
\be
\Gamma^0 = {G_Fm^3\over 8\sqrt{2}\pi}\left[1+\ep(2-\ln m^2) \right].
\ee
The calculation was performed in  $D=4-2\ep$ dimensions and the 
${\cal O}(\ep)$ term is needed for the renormalization of the 
${\cal O}(\alpha_s)$ correction as described below.

The one-loop QCD correction consists of the Born width multiplied by
the wave function renormalization constant $Z_2$ plus contributions of
diagrams (b,c,d).  For $Z_2$ we find ($C_F=4/3$):
\bea
Z_2&=&C_F {g^2\over 16\pi^2} M^{-2\ep}
\left[
-{1\over \ep}
+1 
+\left( 1+{3\over \delta^2}+{8\over \delta^{3/2}} \right)\ln(1-\delta)
+{3\over \delta}   +{8\over \delta^{1/2}}
\right]
\nonumber \\
&=& 
C_F {g^2\over 16\pi^2} M^{-2\ep}
\left[
-{1\over \ep}
-{1\over 2} 
-\sum_{n=1} {\delta^n\over n} 
- 3\sum_{n=1} {\delta^n\over n+2} 
+ {8\over \sqrt{\delta}}\sum_{n=1} {\delta^n\over n+1} 
\right]
\eea

If we pull out a factor $m^3 g_s^2e^2 C_F /(256\pi^3 s_W^2M_W^2)$
the contributions of the diagrams (b,c,d) to the width are
\bea
\Gamma_b &=&
{1\over 2\ep}
-{1\over 2}\ln m^2
-{1\over 2}\ln M^2
+{3\over 4}
- \sum_{n=2} 
\left(
{2(2n^2-1)\over (n^2-1)n^2}  
- {1\over n}\ln \delta\right)
\delta^{n-3/2}
\nonumber \\
&&+3\sum_{n=2} 
{1\over (n+2)(n+1)n(n-1)}
\delta^{n-1},
\nonumber \\
\Gamma_c &=&
 \sum_{n=1} 
\left(
{2n-1 \over 2(n+1)(n+3)}
- {2n\over (n+2)(n+3)}\sqrt{\delta}\right)
\delta^n,
\nonumber \\
\Gamma_d &=&
-{1\over 4\ep}
+{1\over 2}\ln m^2
-{29\over 24}.
\eea

After adding all four contributions and taking $\delta=1$ we find
\bea
\Gamma^{(1)}
&=& 
{C_Fg_s^2 e^2m^3\over s_W^2M_W^2}{1\over 2^9\pi^3} 
\left[
-{13\over 6}
+4 \sum_{n=1} {n^2-n-1\over n(n+1)^2(n+2)} 
\right.
\nonumber\\
&& \left.
- 3 \sum_{n=2} 
{2n^2-3\over (n^2-1)n(n+2)}  
\right]
=
{G_F m^3\over \sqrt{2}}{C_F \alpha_s \over 16\pi^2} 
\left( {5\over 2}-{2\pi^2\over 3}\right).
\eea
The two sums in the above formula correspond to terms odd and even in
$\sqrt{\delta}$, respectively. In obtaining them one has to shift the
summation index so that equal powers of $\delta$ are added together. 
We see that we have  correctly reproduced the result for the one-loop
correction to $t\to bW$.~\cite{jk2}

\section{Summary}
We have reviewed our recent calculations of the
second order corrections to processes involving heavy quarks.  The
techniques we have developed can be applied in many situations where
the radiating particles can be considered non-relativistic.  We have
also demonstrated that some relativistic calculations can also be
performed if sufficiently many expansion terms are available.
Finally, we have discussed some recent results on semileptonic
decays of the heavy quark into a massless quark in the final state.

\section*{Acknowledgments}
We thank N. G. Uraltsev for collaboration on some of the topics
discussed here. 
This work was supported in part by DOE under grant number
DE-AC02-98CH10886, by BMBF under grant number BMBF-057KA92P, and by
Gra\-du\-ier\-ten\-kolleg ``Teilchenphysik'' at the University of Karlsruhe.


\section*{References}
\begin{thebibliography}{99}

\bibitem{BSU} I. Bigi, M. Shifman, and N. Uraltsev,
{\em Ann. Rev. Nucl. Part. Sci.} {\bf 47}, 591 (1997).

\bibitem{VS} 
F. V. Tkachev, {\em Sov. J. Part. Nucl.} {\bf 25}, 649 (1994)
[hep-ph/9701272]. \\
V. A. Smirnov, {\em Mod. Phys. Lett.}  A {\bf 10}, 1485
(1995) [hep-th/9412063].

\bibitem{che81} K.~G. Chetyrkin and F.V. Tkachov,
\Journal{\NPB}{192}{159}{1981}. 

\bibitem{bro91a} D. J. Broadhurst, \Journal{\ZPC}{54}{599}{1992}.

\bibitem{eiko} 
V. A. Smirnov, \Journal{\PLB}{394}{205}{1997}.\\
A. Czarnecki and V. A. Smirnov, \Journal{\PLB}{394}{211}{1997}.

\bibitem{maxtech} A. Czarnecki and K. Melnikov, 
\Journal{\PRD}{56}{7216}{1997}.

\bibitem{Czarnecki:1997hc}
A. Czarnecki and K. Melnikov, \Journal{\PRL}{78}{3630}{1997}.

\bibitem{onehalf}  A. Czarnecki and K. Melnikov,
 hep-ph/9804215.

\bibitem{zeroan}  A. Czarnecki
\Journal{\PRL}{76}{4124}{1996};
A. Czarnecki and K. Melnikov,
\Journal{\NPB}{505}{65}{1997}.

\bibitem{BLM}
S.~J. Brodsky, G.~P. Lepage, and P.~B. Mackenzie, 
\Journal{\PRD}{28}{228}{1983}.

\bibitem{Neubert95beta}
M. Neubert, \Journal{\PLB}{341}{367}{1995}.

\bibitem{Luke:1995}
M. Luke, M.~J. Savage, and M.~B. Wise, {\em Phys. Lett.}  B {\bf 345},  
   301  (1995).

\bibitem{Ball:1995} 
P. Ball and M. Beneke and V. M. Braun, \Journal{\PRD}{52}{3929}{1995}.

\bibitem{BBpriv}
P. Ball and M. Beneke, private communication.

\bibitem{upset} N. G. Uraltsev, {\em Int. J. Mod. Phys.} A {\bf  11},
515 (1996).

\bibitem{BB} M. Beneke and V. M. Braun,
\Journal{\PLB}{348}{513}{1995}. 

\bibitem{CzMeUr1} A. Czarnecki, K. Melnikov and 
N. Uraltsev, \Journal{\PRL}{80}{3189}{1998}. 

\bibitem{Klecture} N. G. Uraltsev, lectures given at 
Intl.~School of Physics ``Enrico Fermi,'' Varenna, Italy, 1997;
hep-ph/9804275. 

\bibitem{CzMeUr} A. Czarnecki, K. Melnikov and 
N. Uraltsev, \Journal{\PRD}{57}{1769}{1998}. 

\bibitem{5infty}  I. Bigi, M. Shifman, N. Uraltsev and A. Vainshtein,
\Journal{\PRD}{56}{4017}{1997}.

\bibitem{topCzMe} A. Czarnecki and K. Melnikov, hep-ph/9806244.

\bibitem{Czarnecki:1996ei} 
 A. Czarnecki and J. H. K\"uhn, \Journal{\PRL}{77}{3955}{1996}.

\bibitem{Czarnecki:1997ac}
 A. Czarnecki and K. Melnikov, \Journal{\PRD}{56}{1638}{1997}. 

\bibitem{jk2}
M. Je{\.z}abek and J.~H. K{\"u}hn, \Journal{\NPB}{314}{1}{1989}.

\end{thebibliography}
\end{document}